\documentclass[useAMS,usenatbib,usegraphicx]{mn2e}
\usepackage{amssymb}
\usepackage{amsmath}
\usepackage{multirow}
\usepackage{hyperref}
\usepackage{times}

\title[Delayed hard X-rays from IRAS 18325$-$5926]
  {X-ray time delays from the Seyfert 2 galaxy IRAS 18325-5926}
\author[A.P. Lobban \& W.N. Alston \& S. Vaughan]
  {A.P.~Lobban$^1$\thanks{e-mail: \href{mailto:al290@le.ac.uk}{al290@le.ac.uk}},
  W.N.~Alston$^2$, S.~Vaughan$^1$ \\
  $^1$University of Leicester, X-Ray and Observational Astronomy Group, Department of Physics and Astronomy, Leicester, LE1 7RH, U.K. \\
  $^2$Institute of Astronomy, Madingley Rd, Cambridge, CB3 0HA, U.K.}
\date{Accepted 2014 September 9 for publication in MNRAS}

\pagerange{\pageref{firstpage}--\pageref{lastpage}} \pubyear{2014}

\def\LaTeX{L\kern-.36em\raise.3ex\hbox{a}\kern-.15em
    T\kern-.1667em\lower.7ex\hbox{E}\kern-.125emX}
\def\iras{IRAS 18325$-$5926}

\def\xmm{{\it XMM-Newton}}

\newcommand{\me}{\mathrm{e}}

\newcommand{\mi}{\mathrm{i}}
\newcommand{\sinc}{\mathrm{sinc}}

\begin{document}

\label{firstpage}

\maketitle

\begin{abstract}

Using new \xmm\ observations we detect hard X-ray time lags in the rapid variability of the Compton-thin Seyfert 2 galaxy \iras. The higher-energy X-ray variations lag behind correlated lower-energy variations by up to $\sim$3\,ks and the magnitude of the lag increases clearly with energy separation between the energy bands. We find that the lag-energy spectrum has a relatively simple $\log(E)$ shape. This is quite different in both shape and magnitude from the lags predicted by simple reflection models, but very similar to the hard X-ray lags often seen in black hole X-ray binaries.  We apply several spectral models to the lag-energy spectrum and rule out simple reflection as an origin for the hard lags.  We find that both propagating fluctuations embedded in the accretion flow and electron scattering from material embedded in or behind a cold absorbing medium offer equally good fits to the observed low-frequency hard X-ray lags and are both consistent with the time-averaged spectrum.  Such models will likely look very different outside of {\it XMM--Newton}'s observable bandpass, paving the way for future studies with {\it NuSTAR}.

\end{abstract}


\section{Introduction} \label{sec:introduction}

Accretion on to black holes powers luminous active galactic nuclei (AGN; $M_{\rm BH} \sim 10^6$\,M$_{\odot}$) and black hole X-ray binaries (XRBs; $M_{\rm BH} \sim 10$\,M$_{\odot}$). Both types of system are powerful sources of X-rays, thought to be generated close to the black hole. The X-ray spectrum of an unobscured AGN is usually dominated by a power-law component thought to be produced in a `corona' of hot electrons by inverse-Compton scattering \citep{HaardtMaraschi93} of ultraviolet (UV) photons emitted from an optically thick, geometrically thin accretion disc \citep{ShakuraSunyaev73}. Illumination of the disc by these X-rays then produces a `Compton reflection' continuum with associated emission lines, the strongest of which is often Fe\,K$\alpha$ fluorescence at $\sim$6.4\,keV \citep{GeorgeFabian91}. X-ray spectroscopy is a powerful tool for investigating the geometry and physical processes involved. But the X-ray source is also, in most cases, rapidly variable, and X-ray timing studies offer alternative insights. The physical interpretation of the complex temporal behaviours so far observed is currently a source of much debate.

In recent years, observations of X-ray time delays (or \emph{lags}) have revealed a range of different phenomena. Delays between variations in different X-ray energy bands -- with the soft X-ray variations preceding the correlated hard X-ray variations, often with the magnitude of the time lag increasing with the separation of the energy bands -- were first seen in XRBs (e.g. Cygnus X-1: \citealt{Miyamoto88, Cui97, Nowak99, KotovChurazovGilfanov01}). The inverse-Compton scattering thought to be responsible for the X-rays will imprint hard X-ray lags; photons that emerge from the corona with higher energies typically have undergone more scatterings, and so have spent more time in the corona. But it was apparent early on that, at least in the XRBs, the magnitude and timescale dependence of the observed lags were not consistent with a compact corona as usually envisaged (\citealt{Miyamoto88, Miyamoto89, Nowak99}). X-ray reflection by the accretion disc may also lead to a delay in the hard X-ray variations (\citealt{KotovChurazovGilfanov01, Poutanen02}). But in general this model did not match the detailed energy dependence of the lags (\citealt{KotovChurazovGilfanov01, Cassatella12}).

The foremost model to explain the existence of low-frequency hard lags is the `propagating fluctuations' model whereby changes in the local mass accretion rate propagate inwards through the accretion disc, powering an extended corona of hot X-ray producing electrons \citep{Lyubarskii97}.  The corona is stratified such that the spectrum produced from the inner regions is harder than from the outer regions.  As fluctuations pass through the emitting region they drive emission with a softer spectrum before driving emission with a harder spectrum, leading to a hard delay, on average.  This model successfully accounts for many of the observed spectral variability properties of Galactic black hole systems \citep{KotovChurazovGilfanov01, ArevaloUttley06} and a natural expectation might be that the same processes dominate in AGN, only scaled to longer timescales.  The model also offers a natural explanation for the energy dependence of the power spectral density (PSD).  The softer X-ray emission is produced at larger radii within the X-ray emitting region, on average, and radial damping of accretion flow variations means that high-frequency variations are suppressed at larger radii, and hence in the softer X-ray emission. The effect is more high frequency power at higher energies.

Hard X-ray lags were first detected in an AGN by \cite{PapadakisNandraKazanas01} and have since been observed in a number of variable, X-ray bright AGN (e.g. \citealt{VaughanFabianNandra03, McHardy04, Arevalo06, McHardy07, EmmanoulopoulosMcHardyPapadakis11, Kara13, Walton13, AlstonVaughanUttley13, AlstonDoneVaughan14, Zoghbi14}). These variations are typically found to occur at low frequencies ($\sim 10^{-5}$--$10^{-4}$\,Hz) in AGN and may be analogous to the hard lags in the XRBs.

But in addition to the hard lags seen at low frequencies, the same AGN often show soft lags at higher frequencies -- where the more rapid soft X-ray variations lag behind the correlated harder X-ray variations (\citealt{Fabian09, DeMarco13}). These soft lags are often explained as arising from the reverberation signal (see \citealt{Uttley14} for a review) as the primary X-ray emission is reflected by material close to the black hole, including an ionized accretion disc capable of producing strong soft X-ray emission (\citealt{Fabian09, ZoghbiUttleyFabian11, Fabian13}).  An alternative proposal is that both the hard and soft lags arise from scattering of the primary X-ray continuum in more distant circumnuclear material tens to hundreds of gravitational radii from the central source (\citealt{Miller10a, Miller10b}).  However, a lot of support has recently built up behind the small-scale reverberation model through the discovery of Fe\,K features in lag-energy spectra (e.g. \citealt{Zoghbi13, Kara13, Kara13b, Kara14, Marinucci14}) with the Fe\,K lag in NGC 4151 displaying intriguing frequency-dependence \citep{Zoghbi12}.  

Typically, Seyfert 1s (which are not heavily obscured and so the nuclear light can be directly observed over the canonical X-ray bandpass) are routinely observed to display rapid X-ray variability and so have been the subject of the majority of X-ray timing studies.  To date, very few detections of X-ray time lags have been made in obscured AGN / Seyfert 2s.  The exceptions are MCG--5-23-16 (a Seyfert 1.9) and NGC 7314 \citep{Zoghbi13, Zoghbi14}.  Here, we report on an $\sim$200\,ks {\it XMM--Newton} observation of \iras, a nearby ($z = 0.01982$; \citealt{Iwasawa95}) X-ray-bright ($F_{\rm 2-10} \sim 3 \times 10^{-11}$\,erg\,cm$^{-2}$\,s$^{-1}$), Compton-thin Seyfert 2 galaxy \citep{deGrijp85} with an observed X-ray luminosity of $L_{\rm 2-10} \sim 10^{43}$\,erg\,s$^{-1}$ \citep{Iwasawa04}.  It is an intriguing, obscured AGN displaying both significant short-term X-ray variability and evidence for a possible broad Fe\,K$\alpha$ emission line ({\citealt{Iwasawa95, LobbanVaughan14}).  In this paper we investigate the X-ray time lags.


\section{\xmm\ data analysis and reduction} \label{sec:data_analysis_and_reduction}

\iras\ was observed twice with \xmm\ \citep{Jansen01} during 2013: on 2013-09-04 (ObsID: 0724820101; rev2516) for 110\,ks and again on 2013-10-15 (ObsID: 0724820201; rev2537) for 107\,ks.  Here we focus on data acquired with the European Photon Imaging Cameras (EPIC) - the pn and the two Metal-Oxide Semiconductor (MOS) detectors -- which were operated using the thin filter and also in small-window mode ($\sim$71 per cent `livetime' for the pn; $\sim$97.5 per cent for the MOS) to reduce the effect of event pile-up \citep{Ballet99, Davis01}.  Version 13.5 of the \xmm\ Scientific Analysis Software (\textsc{sas}) package was used to process all raw data following standard procedures.  

To minimize background contribution, source events were extracted from 20\,arcsec circular regions centred on the source.  Background events were extracted from much larger rectangular regions away from both the central source and other nearby background sources.  The pn (MOS) data were filtered for good X-ray events using the standard {\tt FLAG==0} and {\tt PATTERN $\leq$ 4 (12)} criteria.  The resultant count rates\footnote{The background count rate (for the total band) is $<$1 per cent of the source rate for both observations.} and net exposure times are given in Table~\ref{tab:observation_log} and the broad-band fluxes (averaged over the three EPIC cameras) were found to be $F_{\rm 0.2-10} = 3.55 \times 10^{-11}$\,erg\,cm$^{-2}$\,s$^{-1}$ ($F_{\rm 2-10} = 3.01 \times 10^{-11}$\,erg\,cm$^{-2}$\,s$^{-1}$) and $F_{\rm 0.2-10} = 2.56 \times 10^{-11}$\,erg\,cm$^{-2}$\,s$^{-1}$ ($F_{\rm 2-10} = 2.15 \times 10^{-11}$\,erg\,cm$^{-2}$\,s$^{-1}$) for the rev2516 and rev2537 observations, respectively.

\begin{table}
\centering
\begin{tabular}{l c c c}
\hline
\multirow{2}{*}{Obs\,ID} & \multirow{2}{*}{Camera} & Net exposure & Count rate (0.2--10\,keV) \\ 
& & (ks) & (cts\,s$^{-1}$) \\ [0.5ex]
\hline
\multirow{3}{*}{rev2516} & EPIC-pn & 73 & 5.476\\
& EPIC MOS\,1 & 101 & 1.841 \\
& EPIC MOS\,2 & 100 & 1.940 \\
\hline
\multirow{3}{*}{rev2537} & EPIC-pn & 66 & 4.056 \\
& EPIC MOS\,1 & 91 & 1.387 \\
& EPIC MOS\,2 & 91 & 1.423 \\
\hline
\end{tabular}
\caption{An observation log showing the broad-band count rates and net exposure times and for the three EPIC cameras on-board \xmm\ for each of the two recent observations of \iras.  The lower exposure times of the EPIC-pn camera are due to its operation in small window mode which has a 71 per cent `livetime'.}
\label{tab:observation_log}
\end{table}


\section{Spectral properties} \label{sec:spectral_properties}

We began by modelling the time-averaged broad-band EPIC-pn spectra from 0.3 to 10\,keV to gain insight into some of the fundamental physical components that may be present in the source.  The \textsc{xspec} v12.8.0 software package \citep{Arnaud96} was used for spectral analysis of the background-subtracted spectra.  We binned the spectra using \textsc{specgroup} such that there were $> 25$ counts per bin so as to allow for $\chi^{2}$ minimization and applied the additional criterion that no individual bin be narrower than $1/3$ of the full-width at half-maximum of the detector resolution so as to prevent oversampling.  Confidence intervals are quoted to 90 per cent confidence for one parameter of interest (i.e. $\Delta \chi^{2} = 2.706$) unless stated otherwise.  

We took the approach of fitting the simplest form of the continuum model used to fit previous {\it Suzaku} and \xmm\ observations of \iras, as described in \citet{Tripathi13} and \citet{LobbanVaughan14}.  This is a model of the form \textsc{tbabs}$_{\rm Gal}$ $\times$ [(\textsc{tbabs} $\times$ PL$_{\rm int.}$) + PL$_{\rm scatt.}$ + Gauss].  \textsc{tbabs} models absorption by neutral gas and dust \citep{WilmsAllenMcCray00}, utilizing the photoionization absorption cross-sections of \citet{Verner96}.  The \textsc{tbabs}$_{\rm Gal}$ component accounts for the Galactic hydrogen column, which is fixed at a value of $8.36 \times 10^{20}$\,cm$^{-2}$ based on the measurements of \citet{Kalberla05} at the position of this source and modified to take in account the effect of molecular hydrogen (H$_{\rm 2}$), according to \citet{Willingale13}.  The PL$_{\rm int.}$ component is the intrinsic primary power-law continuum which is absorbed by a column of neutral line-of-sight material using the \textsc{tbabs} code.  Finally, the PL$_{\rm scatt}$ and `Gauss' components refer to a component of scattered continuum emission which dominates in the soft band (modelled with a power law) and an emission line/complex due to Fe which peaks between 6 and 7\,keV, respectively.

The free parameters in the model are the column density ($N_{\rm H}$) of the neutral absorber, the photon index ($\Gamma$) of the primary power-law continuum and the centroid energy ($E_{\rm c}$), intrinsic width ($\sigma$) and normalization of the Gaussian.  The photon index of the scattered power-law component was tied to that of the primary power law while allowing their respective normalizations to also go free.  There was no statistical requirement for a different photon index for the scattered power law as allowing it to vary independently did not significantly improve the fit statistic and the best-fitting value became unconstrained.

The model was applied to the EPIC-pn spectra from both observations and adequately fits the shape of the broad-band continuum returning fit statistics of $\chi^{2} / {\rm d.o.f.}$ (degrees of freedom) $= 2506 / 1930$ and $\chi^{2} / {\rm d.o.f.} = 2528 / 1930$ for the rev2516 and rev2537 spectra, respectively\footnote{We have compared EPIC-pn and MOS spectra and EPIC-pn singles and doubles spectra and are satisfied that our analysis here is not significantly affected by charge transfer inefficiency (CTI) gain issues ({\it XMM--Newton} release note: XMM-CCF-REL-309).}.  The fitted data and their respective residuals are shown in Fig.~\ref{fig:rev2516_rev2537_tbabs_po_po_zga_data_ratio} and the best-fitting values of the various free parameters are detailed in Table~\ref{tab:model_parameters} (note that all fit parameters are given in the rest frame of the galaxy).  The residuals are dominated by the apparent absorption features at $\sim$1.3 and $\sim$1.9\,keV, although the latter feature may be dominated by calibration uncertainties around the Si\,K edge.  On the whole, the best-fitting parameters are similar to those found in previous analyses of this source (e.g. \citealt{Iwasawa04, Tripathi13, LobbanVaughan14}): $N_{\rm H} \sim 10^{22}$\,cm$^{-2}$ and $\Gamma \sim 2$.  The centroid energy of the Fe line is $\sim$6.7\,keV, suggestive of an origin in ionized material and could be indicative of a broad emission line (FWHM $\sim 30\,000$\,km\,s$^{-1}$; equivalent width, EW $\sim 100$\,eV), again consistent with previous observations\footnote{We note that the Fe emission profile in the {\it Suzaku} spectrum acquired in 2006 may be alternatively explained as a blend of narrow lines (see \citealt{LobbanVaughan14}).}.  Finally, the variability between the two observations appears to be simple and is largely dominated by a 50 per cent drop in the normalization of the primary power law between observations, consistent with the variability observed in the {\it Suzaku} data acquired in 2006 \citep{LobbanVaughan14}.

\begin{figure}
\begin{center}
\rotatebox{-90}{\includegraphics[width=6cm]{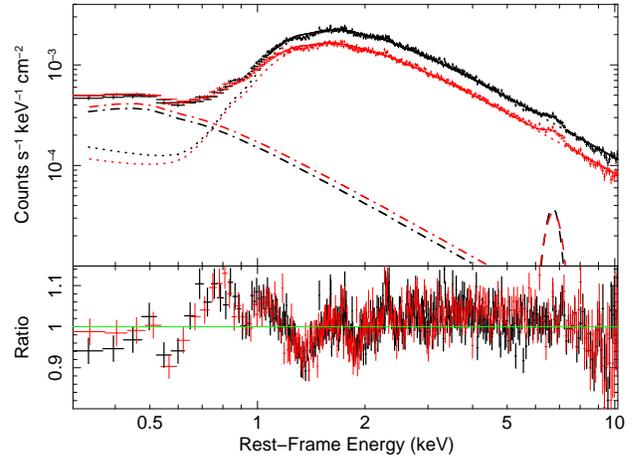}}
\end{center}
\caption{Upper panel: the EPIC-pn spectra from rev2516 (black) and rev2537 (red) \xmm\ observations of \iras\ fitted with a simple model consisting of an absorbed power law, a `scattered' power law dominated in the soft band and a Gaussian modelling the Fe\,K emission complex.  The respective contributions are shown as follows: primary absorbed power law: dotted line; scattered power law: dot-dashed line; Gaussian emission line: dashed line.   For plotting purposes the effective area of the instrument has been divided out.  Lower panel: the ratio of the data-to-model residuals which are dominated by the apparent absorption features at $\sim$1.3 and $\sim$1.9\,keV.}
\label{fig:rev2516_rev2537_tbabs_po_po_zga_data_ratio}
\end{figure}

\begin{table*}
\centering
\begin{tabular}{l c c c c c c c}
\hline
& \textsc{tbabs} & \multicolumn{2}{c}{PL$_{\rm int.}$} & PL$_{\rm scatt.}$ & \multicolumn{3}{c}{Gaussian} \\ 
& $N_{\rm H}$ & $\Gamma$ & Norm. & Norm. & $E_{\rm c}$ & $\sigma$ & Norm. \\
& (cm$^{-2}$) & & (ph\,cm$^{-2}$\,s$^{-1}$) & (ph\,cm$^{-2}$\,s$^{-1}$) & (keV) & (keV) & (ph\,cm$^{-2}$\,s$^{-1}$) \\[0.5ex]
\hline
rev2516 & $1.46^{+0.01}_{-0.01} \times 10^{22}$ & $2.03^{+0.01}_{-0.01}$ & $1.26^{+0.02}_{-0.01} \times 10^{-2}$ & $1.55^{+0.05}_{-0.04} \times 10^{-4}$ & $6.71^{+0.08}_{-0.07}$ & $0.27^{+0.06}_{-0.06}$ & $2.55^{+0.59}_{-0.54} \times 10^{-5}$ \\
rev2537 & $1.29^{+0.01}_{-0.02} \times 10^{22}$ & $1.99^{+0.01}_{-0.01}$ & $0.83^{+0.01}_{-0.02} \times 10^{-2}$ & $1.78^{+0.05}_{-0.04} \times 10^{-4}$ & $6.77^{+0.07}_{-0.07}$ & $0.28^{+0.08}_{-0.06}$ & $2.31^{+0.54}_{-0.50} \times 10^{-5}$ \\
\hline
\end{tabular}
\caption{The best-fitting values of the free parameters in the simple spectral model fitted to the two EPIC-pn spectra of \iras.  The model is of the form \textsc{tbabs}$_{\rm Gal}$ $\times$ [(\textsc{tbabs} $\times$ PL$_{\rm int.}$) + PL$_{\rm scatt.}$ + Gauss] and is described in Section~\ref{sec:spectral_properties}.}
\label{tab:model_parameters}
\end{table*}

Some further structure in the residuals remains (e.g. Fig.~\ref{fig:rev2516_rev2537_tbabs_po_po_zga_data_ratio}; lower panel), largely dominated by an absorption feature at $\sim$1.3\,keV (this was mentioned previously in \citealt{LobbanVaughan14}) -- a more detailed decomposition of the spectrum will be discussed in a forthcoming paper.

Below $\sim$2\,keV, the dominant bright power law begins to become absorbed.  Then, at energies $<0.7$\,keV, the continuum becomes dominated by the smooth scattered power law.  This contrasts with most unobscured AGN which frequently display prominent components of soft excess emission relative to the primary (bright) power law (e.g. \citealt{GierlinskiDone04}, \citealt{Crummy06}, \citealt{ScottStewartMateos12}).  However, it may simply be the case that any component of intrinsic soft excess is diluted by the presence of the scattered power law and/or undetectable due to the large column of absorbing material affecting the spectrum at low energies.

The soft `scattered' power-law component which dominates below $\sim$0.7\,keV appears to be constant in flux over time, displaying no significant variability.  We compared this with the \xmm\ data acquired in 2001 to investigate whether this component was observed to vary on longer timescales.  No useful EPIC-pn data were acquired in 2001 and so we compared the EPIC MOS data so as to minimize any calibration discrepancies between instruments.  By applying the spectral model described above, the flux of this component from 0.2--2\,keV in the MOS\,2 data appears to be constant over this 12-year period: $F_{\rm 0.2-2} = 4.80 \pm 0.28 \times 10^{-13}$\,erg\,cm$^{-2}$\,s$^{-1}$ in 2001 compared with $5.13 \pm 0.39 \times 10^{-13}$\,erg\,cm$^{-2}$\,s$^{-1}$ in 2013, consistent with an origin in material distant from the black hole.  The MOS\,1 data typically predict fluxes $\sim$50 per cent higher $< 2$\,keV in both 2001 and 2013.  Since the MOS\,2 data are more consistent with the pn data, we accept these values and their implication that the soft X-ray flux does not significantly vary.  

\subsection{Reflection models} \label{sec:reflection_models}

In \citet{LobbanVaughan14}, motivated by the shape of the Fe emission complex from 6 to 7\,keV, the broad-band {\it Suzaku} spectrum was modelled with neutral and highly-ionized reflection models.  We took the same approach here, first replacing the Gaussian in the model described in Section~\ref{sec:spectral_properties} with a \textsc{pexmon} \citep{Nandra07}.  This models the reflected continuum from a cold, optically-thick slab when irradiated by a power law while self-consistently modelling emission lines from neutral Fe\,K$\alpha$ fluorescence, Fe\,K$\beta$, Ni\,K$\alpha$ and the Fe\,K$\alpha$ Compton shoulder.  We tied the photon index and normalization to that of the bright, primary power-law component, fixed the cutoff energy at 1\,000\,keV (i.e. consistent with no measurable cut-off), the inclination angle at 60$^{\circ}$ and left the abundances of heavy elements fixed at 1, relative to their interstellar medium (ISM) values \citep{WilmsAllenMcCray00}.  The only free parameter was the reflection scaling factor, $R$.  To estimate the average reflection properties of the source, we applied this model to the summed rev2516+rev2537 pn spectrum.  

Any neutral reflection component appears to be weak with $R = 0.07^{+0.04}_{-0.05}$ (where a value of $R=1$ would correspond to reflection from a standard isotropic source above a disc subtending 2$\pi$\,sr).  The inclusion of the \textsc{pexmon} component only improves the fit by $\Delta \chi^{2} = 8$ and the overall fit is poor ($\chi^{2} / {\rm d.o.f.} = 2861 / 1848$), most likely due to its inability to account for the broadness of the Fe emission complex and the lack of any significant onset of a Compton reflection hump at higher energies.  Allowing the inclination angle to vary did not improve the fit further.  The fit can be improved by $\Delta \chi^{2} = 18$ by allowing the Fe abundance vary, although this would require a best-fitting Fe abundance of $A_{\rm Fe} = 6.4^{+30.7}_{-3.5}$ relative to ISM values.  The free parameters of the power-law and \textsc{tbabs} components did not significantly vary from those quoted in Table~\ref{tab:model_parameters}.

We also tried modelling the spectrum with an ionized reflector, motivated by the apparent emission peak at $\sim$6.7\,keV, suggestive of an origin in ionized material.  We added a \textsc{reflionx} component \citep{RossFabian05} to the model described above and, to account for the apparent broadness of the emission profile, allowed the reflector to be blurred using \textsc{rdblur} within \textsc{xspec}.  \textsc{rdblur} is a convolution kernel based on the \textsc{diskline} model of \citet{Fabian89} taking into account relativistic effects from an accretion disc around a Schwarzschild black hole.  \textsc{reflionx} is designed to model the emergent spectrum when a power law irradiates a photoionized, optically-thick slab of gas. The model assumes a high-energy exponential cut-off, $E_{\rm cut} = 300$\,keV, and uses the abundances of \citet{AndersEbihara82}.  

We based our fit on a more detailed spectral analysis performed on {\it Suzaku} data by \citet{LobbanVaughan14}.  This involved tying the photon index to that of the illuminating power law and initially fixing the Fe abundance at the solar value, leaving the ionization parameter\footnote{The ionization parameter is defined as $\xi = L_{\rm ion} / nR^{2}$ and has units erg\,cm\,s$^{-1}$ where $L_{\rm ion}$ is the ionizing luminosity from 1--1\,000\,Rydberg in units erg\,s$^{-1}$, $n$ is the gas density in cm$^{-3}$ and $R$ is the radius of the absorbing / emitting material from the central source of X-ray in units cm.} and normalization to go free.  The variable parameters of the \textsc{rdblur} code are the inner and outer radii of emission in units of $r_{\rm g}$, the inclination angle of the source to the observer's sightline and the emissivity index, $q$, which defines the radial dependence of emissivity across the disc ($r^{-q}$).  Similar to the {\it Suzaku} spectrum from 2006, the inner and outer radii of emission were fixed at 6\,$r_{\rm g}$ and 500\,$r_{\rm g}$, respectively, and the emissivity index was fixed at $q = 2$.  

Similar to the neutral reflection model, the ionized reflector is driven by the strength of the Fe line.  However, the fit improves by $\Delta \chi^{2} = 80$ compared to the model with just the \textsc{pexmon} component ($\chi^{2} / {\rm d.o.f.} = 2781 / 1845$).  The strength of the neutral \textsc{pexmon} component drops to zero and it becomes no longer statistically required upon the inclusion of the ionized reflector.  The best-fitting value of the ionization parameter of the reflector is $\xi = 550^{+470}_{-90}$\,erg\,cm\,s$^{-1}$, consistent with emission from an ionized disc, while the inclination angle of the \textsc{rdblur} component has a best-fitting value of $\theta = 37^{+5}_{-10}$\,deg.


\section{Variability properties} 
\label{sec:variability_and_time_lags}

Light curves were extracted from the event files using custom \textsc{idl} scripts\footnote{http://www.star.le.ac.uk/~sav2/idl.html}, correcting for exposure losses and interpolating over short telemetry drop outs where necessary. Fig.~\ref{fig:rev2516_rev2537_200-10000_100s_lc} shows the broad-band EPIC-pn lightcurve of \iras\ for both observations in 100\,s time bins. The count rate varies by a factor of $\sim3$ during the observations.  

\begin{figure}
\vspace{-40pt}
\begin{center}
\rotatebox{-270}{\includegraphics[width=6.5cm]{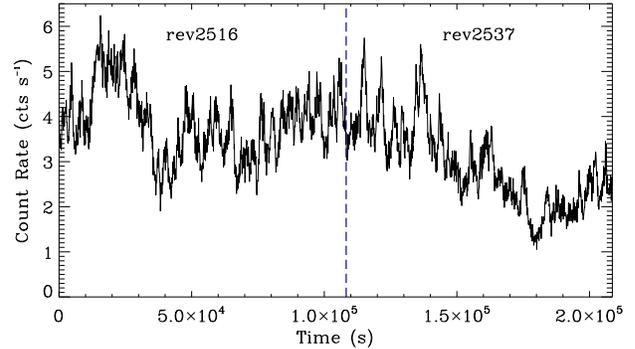}}
\end{center}
\vspace{-20pt}
\caption{The broad-band 0.2--10\,keV \xmm\ EPIC-pn lightcurve of \iras\ in 100\,s bins.  The two observations (rev2516 and rev2537) are not contiguous but are in fact separated by approximately six weeks.  The vertical dashed navy blue line shows where the first observation ends and the second begins.}
\label{fig:rev2516_rev2537_200-10000_100s_lc}
\end{figure}

Fig.~\ref{fig:rev2516_rev2537_pn_rms} shows the fractional rms spectrum (i.e. the rms variability amplitude as a function of energy) computed using the EPIC pn+MOS data and averaged over both orbits (see \citealt{Vaughan03} and references therein). We extracted light curves in $29$ energy bands for each of four 50\,ks segments\footnote{The energy bands are approximately equally spaced in $\log(E)$ with the exception of the three lowest energy bands ($0.2$--$0.4$, $0.4$--$0.8$, $0.8--1.1$~keV) which are made wider to improve the signal-to-noise ratio.} with a timing resolution of $\Delta t = 1$\,ks. We then computed the fractional excess variance ($\sigma_{\rm xs}^2$ / mean$^{2}$) for each band and each segment. Having averaged these values across each segment, we computed the fractional rms by taking the square root of the excess variance.  At energies $>1$\,keV, the fractional rms is approximately constant, but drops sharply at lower energies. This lack of variability most likely arises because we are observing very little direct nuclear light $<0.7$\,keV; the flux is almost entirely scattered light. We therefore exclude the data $<0.7$\,keV from the remainder of the timing analysis.  We also note the apparent drop in variability at $\sim$6--7\,keV, which most likely arises from a component of constant emission from Fe\,K (as in Section~\ref{sec:spectral_properties}).

\begin{figure}
  \begin{center}
    \rotatebox{-270}{\includegraphics[width=6cm]{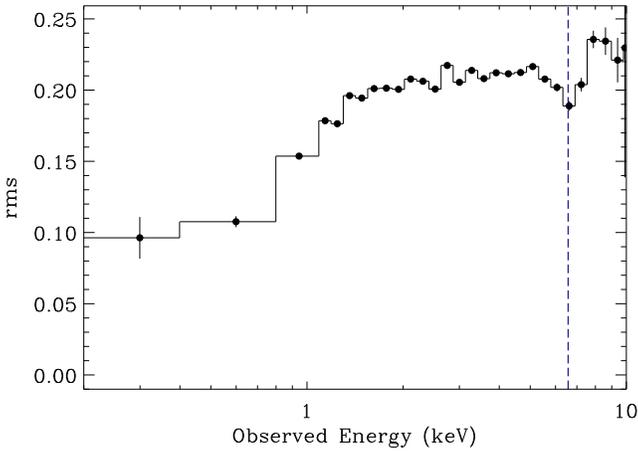}}
  \end{center}
  \caption{The EPIC-pn+MOS rms spectrum of \iras\ averaged over both orbits.  The vertical dashed navy blue line shows the approximate energy of the peak of the Fe\,K emission complex ($\sim$6.7\,keV in the rest frame).}
  \label{fig:rev2516_rev2537_pn_rms}
\end{figure}


\subsection{Time lags as a function of frequency} 
\label{sec:lag-frequency}

We next searched for time delays between X-ray variability at different energies using Fourier methods. We began by comparing variability in two broad energy bands -- `soft' ($0.7$--$1.5$\,keV) and `hard' ($2.5$--$10$\,keV) -- using the cross-spectrum\footnote{These two distinct broad energy bands provide high-quality lag-frequency data.  The finer details of the chosen energy bands do not significantly matter here since we explore the energy dependence of the lags later in Section~\ref{sec:lag-energy} onwards.}. For each of the four $50$ ks segments we computed the Discrete Fourier Transforms, combined these to form auto- and cross-periodograms, and averaged over the four segments to give estimates for the power spectra of the two bands, the coherence and time lags, all as functions of Fourier frequency. We also averaged over contiguous frequency bins each spanning a factor of $\approx 1.7$ in frequency. The method is discussed further in \citet{VaughanNowak97}, \citet{Nowak99}, \citet{Vaughan03} and \citet{Uttley11}. We combined MOS and pn data to maximize the signal-to-noise ratio, and used light curves extracted with $\Delta t = 25$ s bins.

Fig.~\ref{fig:rev2516_rev2537_time_delays_frequency} shows the cross-spectral products.  The upper panel shows the power spectra for the $0.7$--$1.5$\,keV and $2.5$--$10$\,keV energy bands. The middle panel shows the coherence between the two energy bands after Poisson noise correction.  The coherence is a measure of the linear correlation between the two energy bands and is calculated from the magnitude of the cross-periodogram (see \citealt{VaughanNowak97}).  A coherence of 0 means no correlation; a coherence of 1 means the variability in one energy band can be perfectly linearly predicted by the other. The coherence is high ($\sim$0.8--1) at low frequencies (up to $\sim$10$^{-3}$\,Hz), meaning the soft and hard bands are well correlated on long timescales (i.e. $>2$\,ks).  On shorter timescales, the coherence values are not well constrained.

The lower panel of Fig.~\ref{fig:rev2516_rev2537_time_delays_frequency} shows the time lags as a function of frequency.  At high frequencies the two energy bands are consistent with having zero lag.  However, at frequencies $<10^{-4}$\,Hz we detect a significant hard lag (where positive values indicate the hard band lagging behind the soft band) increasing roughly as a power law to a maximum time delay of $\sim$2.5\,ks at $\sim$2 $\times 10^{-5}$\,Hz. There is no indication of the soft lags seen at high frequencies in unabsorbed Seyferts. 

\begin{figure}
\vspace{-20pt}
  \begin{center}
\hspace{-0.5em}
    \rotatebox{-270}{\includegraphics[width=10.2cm]{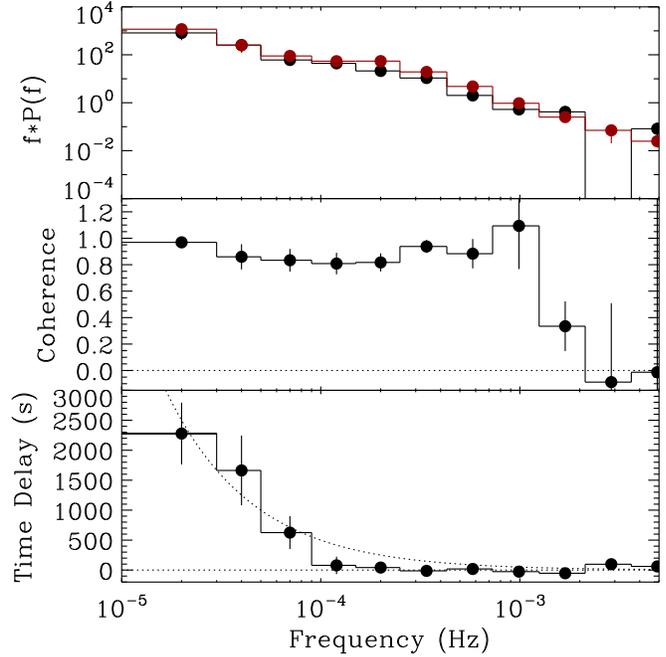}}
  \end{center}
\vspace{-20pt}
  \caption{The cross-spectral products for the soft (0.7--1.5\,keV) and hard (2.5--10\,keV) X-ray bands.  Upper panel: the PSD for the soft (black) and hard (red) bands.  Middle panel: the coherence between the two energy bands.  Lower panel: the time lag between the hard and soft band (a positive value denotes the hard band lagging behind the soft band).  The dotted curve shows a $\tau = 0.05 f^{-1}$ curve; i.e. a lag of $5$\,per cent at each frequency.}
  \label{fig:rev2516_rev2537_time_delays_frequency}
\end{figure}


\subsection{Time lags as a function of energy} 
\label{sec:lag-energy}

Motivated by the detection of a significant hard lag, we investigated the lag as a function of energy.  The lag-energy spectrum is calculated over a given frequency range where a cross-spectral lag is calculated for a series of consecutive energy bands against a standard broad reference band (e.g. \citealt{Uttley11}, \citealt{ZoghbiUttleyFabian11}, \citealt{AlstonDoneVaughan14}).  The choice of reference band produces a systematic offset in the resultant spectrum.  Here, we generated cross-spectral products for 14 logarithmically-spaced energy bands from 0.7 to 10\,keV against a soft reference band from 0.7 to 1.5\,keV\footnote{We also computed lag-energy spectra against a broad reference band consisting of the full 0.7--10\,keV energy range minus the energy band of interest.  The results were consistent.}.  In this instance a positive lag indicates that the given energy band lags behind the soft reference band.  We computed lag-energy spectra over the three lowest frequency bins obtained from our lag-frequency analysis (see Fig.~\ref{fig:rev2516_rev2537_time_delays_frequency}): 1--3 $\times 10^{-5}$, 3--5 $\times 10^{-5}$ and 5--9 $\times 10^{-5}$\,Hz.

The lag-energy spectra over these three frequency ranges are shown in Fig.~\ref{fig:rev2516_rev2537_lag-energy} (panel a)\footnote{Errors on the individual lag estimates in each band were calculated using the standard method (e.g. \citealt{BendatPiersol10}). These are expected to be conservative errors in the sense that they overestimate the scatter in the lags between adjacent energy bins. The reduced scatter results from the fact the light curves involved in the lag estimate for each band are not independent realizations of a random process, but are all highly correlated.}.  At the lowest frequencies, there is a clear correlation between time delay and energy with the time lag scaling linearly with the logarithm of the separation between energy bands.  Very similar behaviour is observed in the XRB systems Cygnus X-1 \citep{Nowak99} and GX 339-4 \citep{Uttley11}.  At higher frequencies ($3$--$5\times 10^{-5}$\,Hz) the magnitude of the lags in any given energy band is smaller (see Fig \ref{fig:rev2516_rev2537_time_delays_frequency}); nevertheless, the same energy dependence is  evident in the frequency range. At higher frequencies the lag is too small (in magnitude) to be clearly detected in any narrow energy band.  Although the observed energy dependence of the time lags is consistent with a number of other AGN (e.g. \citealt{McHardy04}, \citealt{Kara13}), this is one of the clearest detections to date of a log-linear energy dependence.

\begin{figure}
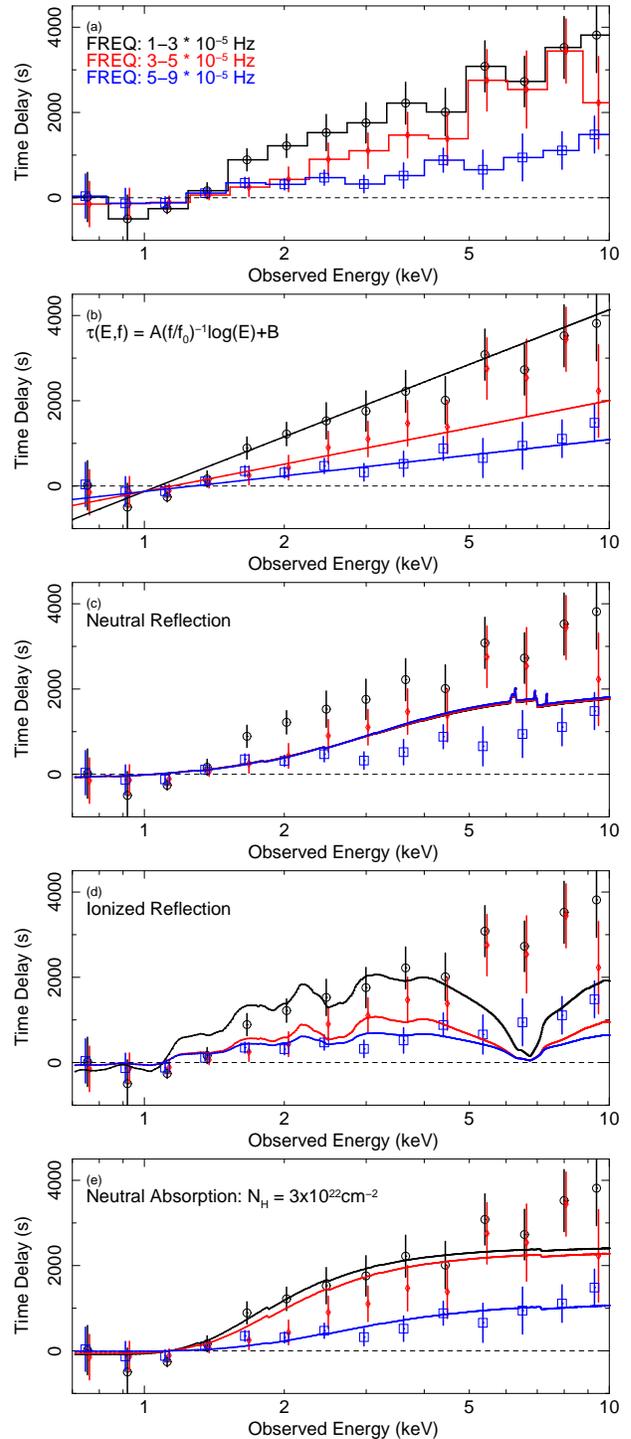

  \begin{center}
    \rotatebox{-90}{\includegraphics[width=3.8cm]{lag-e_ref700-1500.ps}}
    \rotatebox{-90}{\includegraphics[width=3.8cm]{rev2516_rev2537_lag-e_lin-fit_three.ps}}
    \rotatebox{-90}{\includegraphics[width=3.8cm]{rev2516_rev2537_lag-e_pexmon-fit_three.ps}}
    \rotatebox{-90}{\includegraphics[width=3.8cm]{rev2516_rev2537_lag-e_reflionx-fit_three.ps}}
    \rotatebox{-90}{\includegraphics[width=3.8cm]{rev2516_rev2537_lag-e_3e22tbabs-fit_three.ps}}
  \end{center}
  \caption{Panel (a): the lag-energy spectrum of IRAS 18325-5926 against a soft 0.7--1.5\,keV reference band.  The black (circles), red (diamonds) and blue (squares) data points correspond to the 1--3 $\times 10^{-5}$, 3--5 $\times 10^{-5}$ and 5--9 $\times 10^{-5}$\,Hz frequency bands; i.e. the three lowest-frequency bins in Fig.~\ref{fig:rev2516_rev2537_time_delays_frequency}.  The lower panels show the lag-energy spectra when fitted in terms of (b): a simple model of the form: $\tau(E, f) = A (f/f_0)^{-1} \log(E) + B$, (c): neutral reflection, (d) ionized reflection (upon allowing the model to `phase flip') and (e): an absorbed power law with $N_{\rm H} = 3 \times 10^{22}$\,cm$^{-2}$.  High-resolution forms of the fits are shown for display purposes, although the data were fitted with histograms consisting of energy bands matching those in the observed lag-energy spectrum.  The horizontal lines in the lower panels are removed for clarity.  See Sections~\ref{sec:lag-energy} and~\ref{sec:modelling_lags} for details.}
  \label{fig:rev2516_rev2537_lag-energy}
\end{figure}


\section{Modelling the lag-energy data}
\label{sec:modelling_lags}

\subsection{Continuum lags model} \label{sec:continuum_lags_model}

The lag-energy data shown in Fig.~\ref{fig:rev2516_rev2537_lag-energy} (panel a) show an approximately linear dependence of the lag with $\log(E)$.  \citet{KotovChurazovGilfanov01} argue that such an energy dependence may arise from inward propagations of perturbations within the accretion flow and demonstrate this in the case of Cygnus X-1.  In addition, \citet{ArevaloUttley06} demonstrate how a propagating fluctuations model can reproduce the $f^{-1}$ frequency dependence of the lag observed in accreting black hole systems on the basis of various energy-dependent radial emissivities.  We approximated this case by constructing a simple phenomenological model including $f^{-1}$ dependence on frequency and $\log(E)$ dependence on energy:
\begin{equation}
  \tau(E, f) = A (f/f_0)^{-1} \log(E) + B.
\end{equation}
We fixed $f_0=2\times 10^{-5}$ Hz, the central frequency of our lowest frequency bin, and fitted this model to the lag-energy data in the three frequency bands shown in Fig.~\ref{fig:rev2516_rev2537_lag-energy} (panel b). The model has just two free parameters: $A$ and $B$. We found the best fitting model has $A = 4\,276 \pm 283$, $B=-128 \pm 44$\,s and $\chi^2 = 33.8$ with $40$\,d.o.f.; a perfectly acceptable fit.

\subsection{Reprocessing models} \label{sec:reprocessing_models}

As an alternative to the origin of the hard X-ray lags being intrinsic to continuum variations, the lags may also arise from reprocessing of the primary emission by material some distance from the black hole.  As such, we also fitted the shape of the lag-energy data using various standard spectral components, as follows. First, we assume the spectrum comprises two components, a primary (undelayed) spectrum, $p(E)$, and a reprocessed (delayed) spectrum, $r(E)$.  The emission in an energy band $i$, spanning energies ($E_{0,i}, E_{1,i}$), is the sum of the primary and reprocessed spectra integrated over the energy band, $p_i$ and $r_i$, after applying a delay to the reprocessed component.

For simplicity, we assume the time-shifted response is a tophat function (TH) centred at $\tau_0$, with width $\Delta \tau$. Thus the time response function for band $i$ is
\begin{equation}
  \psi_i(t)  =  p_i \delta(t) + \alpha r_i \times TH(t - \tau_0, \Delta \tau), 
\end{equation}
where $\alpha$ defines the strength of the reprocessed component.  From these time domain response functions we can deduce the corresponding frequency transfer functions:
\begin{equation}
\label{eqn:transfers}
  \Psi_i(f)  =  p_i + \alpha r_i \me^{\mi 2 \pi f \tau_0} \sinc(\pi f \Delta \tau).
\end{equation}
Now the phase of this transfer function is:
\begin{eqnarray}
  \phi_i (f) & = & \arg \Psi_i(f) \nonumber \\
             & = & \arctan \left( 
  \frac{R_i \sin (2 \pi f \tau_0 ) \sinc (\pi f \Delta \tau)}{1 + R_i \cos (2 \pi f \tau_0) \sinc (\pi f \Delta \tau)} 
  \right),
\end{eqnarray}
where, for consistency with Uttley et al. (2014; equation 27), we use $R_i = \alpha r_i / p_i$ as the ratio of reprocessed to direct light in each energy channel.  The $\sinc$ terms are a result of the shape of the TH response. The additional $p_i$ term in the denominator is the contribution to the phase from the primary flux, as this affects the real component but not the imaginary component of the complex transfer functions (equation \ref{eqn:transfers}). The $\sin$ and $\cos$ terms give the delay (relative to the primary emission) due to the centroid of the TH. 

The phase difference between two bands, $i$ and $j$, is the difference in their phases, which can be expressed as a time delay between the responses in two bands:
\begin{equation}
  \tau(f,E_j,E_i) = \frac{1}{2 \pi f} (\phi_j (f) - \phi_i (f)).
\end{equation}

In this simple model the energy dependence of $\tau$ is due to the different spectra of primary (direct) and reprocessed (delayed) components. Now, given a specific spectral shape for the primary and reprocessed components, $p(E)$ and $r(E)$, we can express the expected time delay, relative to same fixed reference band, $E_j$, as a function of energy, $E_i$, at a given frequency, $f$, using two free parameters: the central delay, $\tau_0$, and the strength of the reprocessed component, $\alpha$. As this includes both energy and frequency dependence we can simultaneously fit the lag-energy data in the three different frequency bands. 

At a constant $f$, the $\sinc$ terms do not substantially alter the energy dependence, but they do affect the frequency dependence. As our primary concern is in matching the energy dependence of the lags, not in the detailed frequency dependence (which is not well constrained), we have made an additional assumption about the form of the frequency dependence. In particular, we assume $\Delta \tau = 2 \tau_0$; this means the top hat response function starts at $t=0$ and extends to $t=2 \tau_0$ (with a centroid lag of $\tau_0$), as might be expected for a spherical shell type of reprocessor. 

Before presenting the results we first highlight two properties of this model. In the limit of high $\alpha$ (i.e. one band dominated by reprocessed light), and assuming $\tau_0 < 1/(2f)$, the lag-energy spectrum tends to $\tau \approx \tau_0$, constant with energy. The lags effectively `saturate' at $\tau_0$ when the reprocessing dominates. At frequencies above $2/\tau_0$, the sign of the lag in the cross-spectrum `flips', inverting the lag-energy spectrum.  This occurs, at a given Fourier frequency, because the phase difference is constrained to stay within the range [$-\pi$, $+\pi$] (a shift of $+3/4$ wavelengths cannot be distinguished from one of $-1/4$ wavelengths). See \citet{Miller10b} and \citet{ZoghbiUttleyFabian11} for further discussion of this point.

\subsection{Reprocessing model results} \label{sec:reprocessing_model_results}

Using the approach detailed above we first model the lag-energy data in terms of neutral reflection, as per the spectral model described in Section~\ref{sec:reflection_models}.  A neutral-reflection origin to explain the presence of hard lags in AGN has previously discussed and favoured by \citet{Miller10b}.  We assume the primary spectrum, $p(E)$, is a power law with photon index $\Gamma = 2$ (see Table~\ref{tab:model_parameters}) and the reprocessed spectrum, $r(E)$, is purely neutral reflection from a standard optically-thick disc subtending $2\pi$\,sr (\textsc{pexmon}; $R = 1$). Based on these we compute the spectral ratio $r(E)/p(E)$ and optimize the free parameters, $\tau_0$ and $\alpha$, to give the best fit (minimum $\chi^2$).

We find that neutral reflection does not fit the observed lag-energy spectra particularly well, with a best fit of $\chi^{2} / {\rm d.o.f.} = 96.5 / 40$ ($p=1.4 \times 10^{-6}$) and $\tau_0 = 2.2$\,ks.  The strength of the reflection required to account for the observed lag-energy spectrum is unrealistically high with $\alpha = 35.7$.  This means the harder X-ray bands are reflection dominated and the lag saturates at $\tau_0$.  Indeed, the cold reflection model is highly saturated $>5$\,keV resulting in a diluted emission line strength and a flattening of the (normally hard) reflection continuum.  Such strong reflection is clearly inconsistent with the time-averaged spectrum (Section~\ref{sec:reflection_models}).  This fit is plotted in Fig.~\ref{fig:rev2516_rev2537_lag-energy} (panel c).  Note that the high-resolution form of the model is shown in the plot for display purposes -- however, in reality, each fit was performed by integrating over each energy bin such that they are identical to the logarithmically-spaced binning in the lag spectra.

The same approach was also taken in terms of the ionized reflection model described in Section~\ref{sec:reflection_models}.  The motivation for such a model arises from the fact that the Fe emission complex peaks at $\sim$6.7\,keV, which may suggest the presence of a highly ionized accretion disc, as also suggested by \citet{Iwasawa95,Iwasawa04}.  Here, the primary (undelayed) spectrum, $p(E)$, is again a power law with photon index $\Gamma = 2$, while the reprocessed spectrum, $r(E)$, is blurred ionized reflection modelled with \textsc{rdblur} $\times$ \textsc{reflionx} with $\xi \sim 500$\,erg\,cm\,s$^{-1}$, as in Section~\ref{sec:reflection_models}, and again assuming $R=1$\footnote{The \textsc{reflionx} ionized reflection model does not have the reflection scaling factor, $R$, as a free parameter.  Instead, $R$ can be estimated from the ratio of the extrapolated direct power-law flux to the reflected flux. A `standard' reflection value of $R=1$ was estimated by adjusting the \textsc{reflionx} normalization such that this flux ratio was equal to $1$.  Since the \textsc{reflionx} model assumes a high-energy roll-over at 300\,keV, the flux ratio was calculated over the 0.1--200\,keV energy range.}.

We find that the fit is no better than a constant ($\chi^{2} / {\rm d.o.f.} = 327.4 / 40$; i.e. completely unacceptable).  The only way an ionized reflection model can fit the lag-energy spectra is if $\tau_0$ is allowed to roam freely such that `phase flipping' be allowed to occur (e.g. $> 25$\,ks in the case of the $2 \times 10^{-5}$\,Hz frequency band).  In this instance the fit improves to $\chi^{2} / {\rm d.o.f.} = 114.5 / 40$.  This is shown in Fig.~\ref{fig:rev2516_rev2537_lag-energy} (panel d) and has a best-fitting lag of $\tau_0 = 39$\,ks ($\alpha = 5.8$, assuming $R=1$).  An `inverted' reflection spectrum (after `phase-flipping') matches the data better because the lag-energy spectrum is hard (lags increasing in magnitude to higher energies), while the ionized reflection spectrum is quite soft.

Finally, we also fitted the lag-energy spectrum in terms of a scattering model.  Here, we assume a simple case of two different sightlines: one is a direct `undelayed' power law while the `delayed' component is assumed to arise via electron scattering in Thomson-thin material situated either behind or embedded in some cold absorbing material.  Here, the photons that are scattered into our line-of-sight must have travelled a longer path than directly observed photons and so have a greater probability of being absorbed.  The photons that survive scattering/absorption are therefore harder and hence the lagged emission has a harder (more absorbed) spectrum.  This is spectrally akin to a standard partial-covering model and we note that X-ray scattering in terms of time lags has been discussed by \citet{Legg12} (also see \citealt{Miller10a}).  

We tested this scenario using a standard power law as our primary component, $p(E)$, and a power-law modified by neutral absorption using \textsc{tbabs} as our reprocessed component, $r(E)$.  We took the best-fitting values for the power law of $\Gamma = 2$ and normalization $= 1 \times 10^{-2}$\,ph\,cm$^{-2}$\,s$^{1}$ and fitted the lag-energy spectrum stepping through column densities in the range $N_{\rm H} = 10^{22}$--$10^{24}$\,cm$^{-2}$ in steps of $N_{\rm H} = 10^{22}$\,cm$^{-2}$.  A column density of $N_{\rm H} = 3 \times 10^{22}$\,cm$^{-2}$ was found to give the best fit with $\chi^{2} / {\rm d.o.f.} = 38.1 / 40$ ($p = 0.56$), $\alpha = 0.67$ and a best-fitting lag of $\tau_0 = 6.4$\,ks.  We note that, from a  $N_{\rm H} = 3 \times 10^{22}$\,cm$^{-2}$ medium, one may expect fluorescence from a variety of atomic transitions, which may be detectable if the covering fraction is significant.  Such lines may potentially make our fit worse although it is difficult to predict how strong such features would be and it is unclear as to whether the resolution of our observed lag-energy spectrum would be high enough to detect them.  Nevertheless, our primary focus is to fit the smooth continuum-form of the lags.  This fit is shown by the solid curves in Fig.~\ref{fig:rev2516_rev2537_lag-energy} (panel e).  We also note the fits at either end of the range of column densities explored: $N_{\rm H} = 10^{22}$ ($\chi^{2} / {\rm d.o.f.} = 96.5 / 40$) and $N_{\rm H} = 10^{24}$\,cm$^{-2}$ ($\chi^{2} / {\rm d.o.f.} = 160 / 40$), respectively.

While the low-frequency lag-energy spectrum in IRAS 18325-5926 can be fitted well in terms of a neutral absorbed power law with a best-fitting column density of $N_{\rm H} = 3 \times 10^{22}$\,cm$^{-2}$, we tested to see if such a `delayed' component could be consistent with the time-averaged spectrum, adding together with the `direct' component to form the observed spectrum.  We took the baseline spectral model described in Section~\ref{sec:spectral_properties} and included an additional `delayed' absorbed power law with the column density fixed at $N_{\rm H} = 3 \times 10^{22}$\,cm$^{-2}$ and the photon index tied to that of the other power law components.  The form of the model was: \textsc{tbabs}$_{\rm Gal}$ $\times$ [(\textsc{tbabs} $\times$ PL$_{\rm int.}$) + (\textsc{tbabs}$^{\rm 3e22}$ $\times$ PL$_{\rm delayed}$) + PL$_{\rm scatt.}$ + Gauss].  The inclusion of this component significantly improves the fit to the time-averaged spectrum ($\Delta \chi^{2} \sim 400$) from 0.3 to 10\,keV.  The best-fitting photon index, however, increases to $\Gamma \sim 2.2$ and the normalization of the PL$_{\rm delayed}$ component is found to be $\sim$50 per cent of that of the brighter `direct' PL$_{\rm int.}$ component.


\section{Discussion}
\label{sec:discussion}

We have presented an analysis of the short-timescale variability of IRAS 18325-5926 through two $\sim$100\,ks {\it XMM--Newton} observations and detected a hard lag, with the higher-energy variations lagging behind correlated soft variations by up to $\sim$3\,ks.  There is a strong energy-dependence to the lag, the magnitude of which increases approximately linearly with the log of the separation of the energy bands.

Several physical models have been proposed to explain the hard lags in accreting black hole systems.  Reflection of the primary power law (most likely by the accretion disc) could result in a delay in hard X-ray variability (e.g. \citealt{KotovChurazovGilfanov01, Poutanen02}) if the reflection spectrum is hard (as expected for a near-neutral reflector).  Reflection models have been invoked to explain the observed lags in a variety of AGN (e.g. \citealt{Fabian09, ZoghbiUttleyFabian11, Fabian13}) although in these cases the softest band was also delayed and the reflector assumed to be highly ionized.  

We tested the possibility of a reprocessing-origin for the lags in IRAS 18325-5926 by modelling their energy dependence in terms of the primary components that may build up the spectrum of a typical Seyfert galaxy: neutral reflection, ionized reflection and a neutral absorbed power law (Section~\ref{sec:modelling_lags}).  We found that a neutral reflector can recreate the shape of the low-frequency lag-energy spectrum only if the strength of the reflector is unrealistically high.  The reflection strength required to reproduce the lag spectrum grossly overpredicts the strength of the iron line and reflection features seen in the time-averaged spectrum (e.g. Section~\ref{sec:reflection_models}; \citealt{LobbanVaughan14}).  We therefore conclude that neutral reflection cannot be the origin of the low-frequency hard lags observed here.  We reach the same conclusion for ionized reflection, which is spectrally too soft (and weak) to account for the hard shape of the lag-energy spectrum, resulting in a very poor fit.  

\citet{Zoghbi14} analysed recent {\it Suzaku}, {\it XMM--Newton} and {\it NuStar} data for the AGN MCG--5-23-16 and also find a hard shape to their observed lag-energy spectrum.  They conclude that the lags are dominated by relativistic reverberation from material a few tens of gravitational radii from the black hole.  This is primarily due to the non-smooth shape of the spectrum which manifests itself in the form of a peak at $\sim$6\,keV and a possible down-turn at $\sim$30\,keV which they associate with Fe\,K$\alpha$ emission and the Compton reflection hump, respectively.  However, those reverberation lags exist at relatively higher frequencies than those probed in this paper and it is conceivable that hard `continuum' lags would also be observed in MCG--5-23-16 if one could probe the lower-frequency domain.  As such, it is plausible that both processes are at work but dominating on different timescales and possibly differing in strength between sources.

An alternative reprocessing model, however, has the lag produced by electron scattering in Thomson-thin, neutral material at some distance from the central source of X-rays.   A power law absorbed by neutral material provides a much better fit than reflection with $\chi^{2} / {\rm d.o.f.} = 38.1 / 40$ for $N_{\rm H} = 3 \times 10^{22}$\,cm$^{-2}$.  Such a lag-energy spectrum could be produced by simple X-ray scattering occurring in a medium behind or embedded in a cold absorbing medium.  X-ray scattering in the context of time lags has been discussed by \citet{Miller10a} and \citet{Legg12}.  Spectrally, this is the same as the standard partial-covering model (also see \citealt{MillerTurner13} for further discussion on absorption and scattering cloud models).  Intriguingly, fitting the `delayed' absorbed power law ($N_{\rm H} = 3 \times 10^{22}$\,cm$^{-2}$) to the time-averaged spectrum requires that the `delayed' component be a factor of $\sim$2 weaker than the `direct' component, perhaps consistent with the ratio of reprocessed to direct emission obtained from the lag-energy fitting ($\alpha = 0.67$).  

However, the energy dependence of the PSD is a challenge for all the reprocessing models (see Fig.~\ref{fig:rev2516_rev2537_time_delays_frequency}; upper panel).  If the hard lags are due to reprocessing of a hard spectrum with a long delay (as in the reflection and absorbed power law models) this should dampen the high-frequency variations in harder energy bands.  This is because the variations in the reprocessed light are both delayed and smoothed with the smoothing expected to steepen the PSD on timescales $\leq \tau_0$.  However, the higher-energy PSD in Fig.~\ref{fig:rev2516_rev2537_time_delays_frequency} shows no significant steepening towards higher frequencies.  Additionally, the rms spectrum shown in Fig.~\ref{fig:rev2516_rev2537_pn_rms} is flat, which also suggests that there is no suppression of high-frequency variability at higher energies, where reprocessing might be expected to dominate.

Our findings, however, are similar to those of \citet{KotovChurazovGilfanov01} who studied the logarithmic energy dependence of the lags of the XRB Cygnus X-1.  They demonstrated that the expected lag-energy spectrum from a standard extended reflector did not match the observed spectrum, primarily due to the suppression of the lags at $\sim$6.4\,keV, where a more prominent feature due to Fe\,K$\alpha$ fluorescence was expected.  Similarly, we also exclude a reflection origin for the hard lags in IRAS 18325-5926 due to the inability of reflection models to match the observed smooth lag-energy spectrum.   

The favoured model to explain the low-frequency energy dependence of hard lags in XRBs is the `propagating fluctuations' model, based on the model proposed by \citet{Lyubarskii97}, whereby inwardly propagating accretion rate fluctuations (at the radial drift velocity) generated over a wide range of radii in the accretion flow modulate each other. The X-ray emission, from the inner regions, is modulated in turn. An extended emission region. with softer emission produced at larger radii than harder spectra, can produce a net lag as the harder emission responds (on average) later to the inward moving fluctuations.  The similarity of the lags observed in AGN suggests that a similar mechanism may be also at work in larger accreting systems.  \citet{ArevaloUttley06} demonstrate how, for a standard accretion disc, the model can reproduce the observed $\sim$1$/f$ frequency-dependence of time lags between energy bands and their associated amplitudes by invoking different radial emissivities with some given energy dependence.

\citet{KotovChurazovGilfanov01} showed that the logarithmic energy dependence of the hard lag in Cygnus X-1 is compatible with a model based on propagating fluctuations within the accretion flow.  Their observed logarithmic energy-dependence arises directly from the assumption that each locally-emitted spectrum is a power law with the photon index decreasing at smaller radii from the black hole.  In Section~\ref{sec:modelling_lags}, a $\log(E)$ model was found to give a good fit to the lag-energy spectrum of IRAS 18325-5926 with $\chi^{2} / {\rm d.o.f.} = 33.8 / 40$.  As such, the $\log(E)$ shape of the lag-energy spectrum may be consistent with the propagating fluctuations model.  Intriguingly, the low-frequency hard lags observed here and in NGC 6814 \citep{Walton13} are somewhat similar to those observed in many other variable AGN (e.g. \citealt{McHardy04}; \citealt{Fabian13}), despite those AGN typically displaying more `standard' soft X-ray spectra (i.e. with strong soft excesses).  As such, this may lend credence to the hypothesis of an origin in the primary power-law continuum for the low-frequency hard lags.

Studies of low-frequency X-ray lags are important as they may be a ubiquitous feature in (radio-quiet) AGN, whereas the more hotly debated soft lags may not be.  As a common feature in accreting black hole systems of all sizes (XRBs through to AGN), hard lags are most likely carrying valuable information about the structure of inner accretion flows around black holes.  It is also crucial to understand the details of hard lags if we are to disentangle them from high-frequency soft lags and obtain unambiguous answers.   

We currently have good quality data from 1 to 10\,keV with {\it XMM--Newton}.  However, the models that work well [i.e. $\log(E)$ / neutral absorbed power law] make very different predictions outside our observed frequency and energy ranges.  At lower frequencies, the propagating fluctuations model predicts that the lag continues to rise while the absorbed power law has a fixed $\sim$6\,ks lag which should remain constant.  Meanwhile, at higher energies the absorbed power-law model should also show a constant lag-energy spectrum with the relative strength of absorbed and unabsorbed flux becoming invariant above a few keV while saturating to $0$ at the softest energies.  As such, it is hoped that they may be distinguished by extending both the energy range and/or frequency range such as through observations with {\it NuSTAR} \citep{Harrison13}, extending the bandpass up to $\sim$80\,keV, but also with {\it XMM--Newton} if given less absorbed AGN.


\section*{Acknowledgements}

This research has made use of the NASA Astronomical Data System (ADS), the NASA Extragalactic Database (NED) and data obtained from the \xmm\ satellite, an ESA science mission with instruments and contributions directly funded by ESA Member States and the USA (NASA). AL and SV acknowledge support from the UK STFC research council.  WNA acknowledges support from the European Union Seventh Framework Programme (FP7/2007-2013) under grant agreement n.312789, StrongGravity.  We would like to thank Phil Uttley for useful and constructive comments and discussion and we also wish to thank our anonymous referee for a careful and thorough review of the draft.


\label{lastpage}

\end{document}